\documentclass[aps,prb,amsmath,amssymb,twocolumn]{revtex4}
\usepackage{graphicx}
\usepackage{bm}
\usepackage{bbold}

\def\bk{ \bm{k} }

\def\bgam{ \bm{\gamma} }

\def\sgn{\, \mathrm{sgn}\, }

\def\diag{\,\mathrm{diag}\,}

\begin{document}
\title{Noncentrosymmetric superconductors in one dimension}

\author{K. V. Samokhin\footnote{e-mail: kirill.samokhin@brocku.ca}}

\affiliation{Department of Physics, Brock University, St. Catharines, Ontario L2S 3A1, Canada}
\date{\today}

\begin{abstract}
We study the fermionic boundary modes (Andreev bound states) in a time-reversal invariant one-dimensional superconductor. In the presence of a substrate, spatial inversion symmetry is broken and the electronic properties 
are strongly affected by an antisymmetric spin-orbit coupling. We assume an arbitrary even number of nondegenerate bands crossing the Fermi level. We show that there is only one possible pairing symmetry in one dimension,
an analog of $s$-wave pairing. The zero-energy Andreev bound states are present if the sign of the gap function in an odd number of bands is different from all other bands.
 
\end{abstract}

\maketitle

\section{Introduction}
\label{sec: Intro}

One-dimensional (1D) topological superconductors have recently attracted a lot of attention, primarily in the context of searching for topologically-driven physics, in particular, the Majorana quasiparticles, 
in condensed matter systems.\cite{MFs-review,top-SC} Motivated by the observation that a simple 1D model of spinless $p$-wave pairing (the Kitaev chain) can support zero-energy Majorana boundary states,\cite{Kit01} a number 
of proposals for engineering an effective $p$-wave superconductor have been put forward.
For instance, it was shown theoretically in Ref. \onlinecite{TRB-1D} that one can create Majorana states at the ends of a semiconducting wire
which is proximity-coupled to an ordinary $s$-wave superconductor, by applying a sufficienly strong magnetic field. Experimental signatures of such boundary states would include zero-bias peaks in the tunneling conductance, 
which have been reported for InSb nanowires in Refs. \onlinecite{InSb-wire-1} and \onlinecite{InSb-wire-2}. There are other ways to create a time reversal (TR) symmetry-breaking topological superconductor with 
Majorana quasiparticles, e.g., in a chain of Fe atoms on a superconducting Pb substrate, as discussed in Ref. \onlinecite{Fe-chain}. 
Various models of TR invariant 1D superconductors, both with and without topologically nontrivial phases, have also been studied.\cite{TRI-1D}
   
The majority of proposals for topological states of matter rely on the Rashba spin-orbit (SO) coupling, which requires the absence of inversion symmetry in the system,
see Refs. \onlinecite{Rashba-model}, \onlinecite{Manchon15}, and the references therein.  
The SO coupling in a noncentrosymmetric crystal lifts the spin degeneracy of the electron states, producing nondegenerate Bloch bands labeled by ``helicity'', with
wave functions characterized by a complex spin structure and a nontrivial momentum-space topology. Its profound consequences for superconductivity in three-dimensional (3D) and two-dimensional (2D) materials 
have been extensively studied in the last decade, see Refs. \onlinecite{NCSC-book}, \onlinecite{Kneid15}, and \onlinecite{Smid16} for reviews.  

In this paper, we study superconductivity in a metallic quantum wire on a substrate, in the absence of an external magnetic field or any other TR symmetry breaking mechanism. The pairing interaction can be either intrinsic 
to the wire or extrinsic, i.e., induced by the substrate. In contrast to the previous works, we do not limit ourselves to any particular models of the SO coupling and the superconducting pairing. 
Our goal is to find the spectrum of the fermionic states localized near the end of the wire, known as the Andreev bound states (ABS), in the most general setup, assuming only the point-group and 
TR invariance. 

Due to the lack of inversion symmetry, the electrons in the wire are affected by the antisymmetric SO coupling and form nondegenerate 1D bands.  
We consider an arbitrary number $N$ of such bands crossing the Fermi level ($N$ can be shown to be even), and present a complete symmetry analysis of the SO coupling and the superconducting gap structure, 
in Secs. \ref{sec: electron bands} and \ref{sec: SC}, respectively. In Secs. \ref{sec: ABS} and \ref{sec: TRI}, we calculate the ABS spectrum using the semiclassical approach, 
with the boundary conditions formulated in terms of the scattering matrix.     
Throughout the paper we use the units in which $\hbar=k_B=1$, neglecting, in particular, the difference between the quasiparticle momentum and wave vector.

\section{Spin-orbit coupling and electron bands in 1D}
\label{sec: electron bands}

We consider a quasi-1D electron gas, which can be created, for instance, by applying a gate voltage in the $y$ direction to 2D electrons on a $xy$-plane substrate. 
Thus, the 3D potential $U(x,y,z)$ affecting the electrons is assumed to be periodic or constant in $x$ direction, but confining in both $y$ and $z$ directions. This system is TR invariant in the normal state 
but lacks an inversion center, because the substrate breaks the $z\to-z$ mirror reflection symmetry. The momentum space is one-dimensional, labelled by the wave vector $\bk=k_x\hat{\bm{x}}$,
which takes values in the first Brillouin zone (BZ): $-\pi/d<k_x\leq\pi/d$, where $d$ is the period of the 1D lattice. 

In addition to the TR symmetry, the system is also invariant under the crystallographic point operations.
In contrast to higher-dimensional noncentrosymmetric systems,\cite{Sam15} the list of possible symmetries in quasi-1D is very limited, 
generated by just two reflections $\sigma_x$ and $\sigma_y$. Their action is defined 
in the standard fashion, as $\sigma_xU(x,y,z)=U(-x,y,z)$ and $\sigma_yU(x,y,z)=U(x,-y,z)$, while their combination is equivalent to a $\pi$ rotation about the $z$ axis: $\sigma_x\sigma_yU(x,y,z)=U(-x,-y,z)$. 
It is easy to see that there are only five quasi-1D point groups, all Abelian, which are listed in the first column of Table \ref{1D-groups} (Ref. \onlinecite{1D}). 
The groups $\mathbf{D}_x$, $\mathbf{D}_y$, and $\mathbf{C}_2$ are isomorphic to the cyclic group $\mathbb{Z}_2$, while $\mathbf{V}$ is isomorphic to the Klein four-group $\mathbb{Z}_2\times \mathbb{Z}_2$.

Due to the absence of inversion symmetry, the electron bands, which we label by the index $n$, are nondegenerate. The Bloch states with opposite momenta form a Kramers pair, 
therefore the band dispersions are even functions of momentum:
\begin{equation}
\label{xi-even}
  \xi_n(k_x)=\xi_n(-k_x).
\end{equation}
The energies are counted from the chemical potential and the difference between the latter and the Fermi energy $\epsilon_F$ is neglected. Since the bands are nondegenerate, the TR action on the Bloch states can be 
written as\cite{t-factor} 
\begin{equation}
\label{t_n}
  K|k_x,n\rangle=t_n(k_x)|-k_x,n\rangle,
\end{equation}
where $t_n(k_x)$ is a phase factor. Recall that the TR operator for spin-1/2 particles is $K=i\hat\sigma_2K_0$, where $\hat\sigma_2$ is the Pauli matrix and $K_0$ is complex conjugation. Since $K^2=-1$ and, 
on the other hand, $K^2|k_x,n\rangle=t^*_n(k_x)t_n(-k_x)|k_x,n\rangle$, we obtain:
\begin{equation}
\label{t-n-odd}
  t_n(-k_x)=-t_n(k_x).
\end{equation}
Note that $t_n(k_x)$ is not gauge invariant: under a rotation of the phases of the band states, $|k_x,n\rangle\to e^{i\theta_n(k_x)}|k_x,n\rangle$, one has 
\begin{equation}
\label{t-transformation}
  t_n(k_x)\to e^{-i[\theta_n(k_x)+\theta_n(-k_x)]}t_n(k_x).
\end{equation}
Therefore, the TR phase factors cannot appear in any observable quantity.  

To illustrate the pecularities of the electronic band structure in noncentrosymmetric 1D systems with the SO coupling, one can use the following Hamiltonian:
\begin{equation}
\label{H-Rashba}
    \hat H_0=\sum\limits_{k_x}\sum_{\alpha,\beta=\uparrow,\downarrow}\left[\epsilon_0(k_x)\delta_{\alpha\beta}+\bgam(k_x)\bm{\sigma}_{\alpha\beta}\right]\hat a^\dagger_{k_x\alpha}\hat a_{k_x\beta}.
\end{equation}
This is the 1D version of the well-known Rashba model, see Refs. \onlinecite{Rashba-model}, \onlinecite{Manchon15}, and the references therein. 
The first term, with $\epsilon_0(k_x)=\epsilon_0(-k_x)$, describes a twofold degenerate band in the absence of antisymmetric SO coupling. While the momentum space is 1D, the spin space is still 3D, so that  
the antisymmetric SO coupling is described by the 3D pseudovector $\bgam(k_x)$, which is real, odd in $k_x$ due to the TR symmetry, periodic in the reciprocal space,
and also invariant under the point group operations, that is $g\bgam(g^{-1}k_x)=\bgam(k_x)$ for all elements $g$ of the point group.
One has to remember that mirror reflections act differently on polar vectors, such as $\bk=k_x\hat{\bm{x}}$ and pseudovectors, such as $\bgam$. We have $\sigma_x k_x=-k_x$ and $\sigma_y k_x=k_x$, but
$\sigma_x(\gamma_x,\gamma_y,\gamma_z)=(\gamma_x,-\gamma_y,-\gamma_z)$ and $\sigma_y(\gamma_x,\gamma_y,\gamma_z)=(-\gamma_x,\gamma_y,-\gamma_z)$. In this way, one obtains that the simplest expression for
the Rashba SO coupling in 1D, taking into account the lattice periodicity, is
\begin{equation}
\label{gamma-Rashba}
  \bgam(k_x)=\bm{a}\sin(k_xd),
\end{equation}
for all five point groups, with $\bm{a}$ given in the second column of Table \ref{1D-groups}. The Rashba model (\ref{H-Rashba}) can be derived from a more general Hamiltonian of 1D electrons in a noncentrosymmetric 
crystal lattice with the SO coupling, see Appendix \ref{sec: ASOC-1D}. 

The Rashba SO coupling results in the lifting of the pseudospin band degeneracy almost everywhere in the BZ. Diagonalizing Eq. (\ref{H-Rashba}), we obtain:
\begin{equation}
\label{Rashba-bands}
   \xi_\lambda(k_x)=\epsilon_0(k_x)+\lambda|\bgam(k_x)|,
\end{equation}
where the band index $\lambda=\pm$ is called the helicity. The corresponding eigenstates can be chosen, for instance, in the following form:
\begin{equation}
\label{Rashba-eigenstates}
  |k_x,\lambda\rangle=\frac{1}{\sqrt{2}}\left(\begin{array}{c}
                                        \sqrt{1+\lambda\sgn k_x\dfrac{a_z}{|\bm{a}|}} \\
					\lambda\sgn k_x\dfrac{a_x+ia_y}{\sqrt{a_x^2+a_y^2}}\sqrt{1-\lambda\sgn k_x\dfrac{a_z}{|\bm{a}|}}
                                        \end{array}\right).
\end{equation}
Substituting these expressions in Eq. (\ref{t_n}), we obtain the TR phase factors in the Rashba model:
$$
  t_\lambda(k_x)=\lambda\frac{a_x-ia_y}{\sqrt{a_x^2+a_y^2}}\sgn k_x.
$$
Note that one can make $t_\lambda(k_x)$ real and equal to $\sgn k_x$ by a suitable gauge transformation of the eigenstates, see Eq. (\ref{t-transformation}). 
The bands (\ref{Rashba-bands}) are nondegenerate at all $k_x$, except the TR invariant points $k_x=0$ and $k_x=\pi/d$, where the Rashba SO coupling (\ref{gamma-Rashba}) vanishes and neither the helicity eigenstates
nor the TR phase factors are well defined.

\begin{table}
\caption{The list of quasi-1D point groups (the first column) and the corresponding Rashba SO coupling vectors (the second column), $a_{1,2,3}$ are
real parameters.}
\begin{tabular}{|c|c|c|}
    \hline
    $\mathbb{G}$ & $\bm{a}$ \\ \hline
    $\mathbf{C}_1=\{E\}$    &  $\ (a_1,a_2,a_3)\ $ \\ \hline
    $\mathbf{D}_x=\{E,\sigma_x\}$   & $(0,a_2,a_3)$ \\ \hline
    $\mathbf{D}_y=\{E,\sigma_y\}$   & $(0,a_2,0)$ \\ \hline
    $\mathbf{C}_2=\{E,\sigma_x\sigma_y\}$  & $(a_1,a_2,0)$  \\ \hline
    $\ \mathbf{V}=\{E,\sigma_x,\sigma_y,\sigma_x\sigma_y\}\ $  & $(0,a_2,0)$ \\ \hline 
\end{tabular}
\label{1D-groups}
\end{table}

\section{Superconductivity in nondegenerate bands}
\label{sec: SC}

\begin{figure}
\includegraphics[width=8cm]{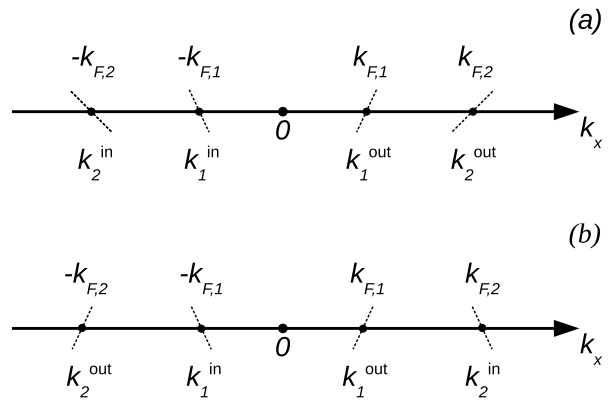}
\caption{The 1D ``Fermi surface'' and the incident and reflected wave vectors for $N=2$. The dashed lines show the band dispersion slopes near the Fermi points, for (a) the Fermi group velocities 
$v_{1,+}$ and $v_{2,+}$ having the same (positive) sign, and (b) $v_{1,+}$ and $v_{2,+}$ having opposite signs.}
\label{fig: Fermi points}
\end{figure}

We assume that there are $N$ nondegenerate bands crossing the Fermi level and participating in superconductivity. According to Appendix \ref{sec: ASOC-1D}, $N$ is an even number. 
The 1D ``Fermi surface'' is given by a set of $2N$ Fermi wave vectors $\pm k_{F,n}$, which are the roots of the equations $\xi_n(k_x)=0$, as illustrated in Fig. \ref{fig: Fermi points}. 
It is straightforward to extend our results to the case of multiple pairs of the Fermi points in the same band. 
The mean-field analysis of superconductivity developed below does not rely on any specific pairing mechanism, the only assumption being that the SO band splitting is large enough
to suppress the pairing of electrons from different bands.

The Hamiltonian is constructed using the basis of the exact band states $|k_x,n\rangle$, which include all effects of the lattice potential and the SO coupling. In the mean-field approximation we have 
\begin{eqnarray}
\label{H-MF}
     && \hat H=\sum_{k_x,n}\xi_n(k_x)\hat c^\dagger_{k_x,n}\hat c_{k_x,n}\nonumber\\
     &&+\frac{1}{2}\sum_{k_x,n}\left[\Delta_n(k_x)\hat c^\dagger_{k_x,n}\hat{\tilde c}^\dagger_{k_x,n}+\Delta^*_n(k_x)\hat{\tilde c}_{k_x,n}\hat c_{k_x,n}\right].
\end{eqnarray}
The first term here describes noninteracting quasiparticles in the nondegenerate bands, while the second term represents the intraband pairing between the states $|k_x,n\rangle$ and $K|k_x,n\rangle$. 
The electron creation operator in the time-reversed state $K|k_x,n\rangle$ is defined as
\begin{equation}
\label{c-tilde c}
  \hat{\tilde c}^\dagger_{k_x,n}\equiv K\hat c^\dagger_{k_x,n}K^{-1}=t_n(k_x)\hat c^\dagger_{-k_x,n},
\end{equation}
where we used Eq. (\ref{t_n}) (note that $K\hat{\tilde c}^\dagger_{k_x,n}K^{-1}=-\hat c^\dagger_{k_x,n}$). Substituting Eq. (\ref{c-tilde c}) in $\hat H$ and using 
the anticommutation of the fermionic creation and annihilation operators, we obtain:
\begin{equation}
\label{Delta-even}
  \Delta_n(k_x)=\Delta_n(-k_x),
\end{equation}
therefore the pairing in the band representation is necessarily even in momentum.

The gap functions $\Delta_n(k_x)$ have simple transformation properties under the point group and TR operations.
Indeed, consider the action of an element $g$ of the point group on the Bloch states: $g|k_x,n\rangle=e^{i\varphi_{k_x,n}(g)}|gk_x,n\rangle$, 
where $\varphi_{k_x,n}(g)$ is a phase. It is straightforward to show that, due to the commutation of $g$ and $K$, we have 
$g(\hat c^\dagger_{k_x,n}\hat{\tilde c}^\dagger_{k_x,n})g^{-1}=\hat c^\dagger_{gk_x,n}\hat{\tilde c}^\dagger_{gk_x,n}$, therefore 
the gap function transforms as a scalar field, i.e., $\Delta_n(k_x)\to\Delta_n(g^{-1}k_x)$.
As a consequence, the Cooper pairing in 1D is always conventional (``$s$-wave'') in the sense that, due to the property (\ref{Delta-even}), 
the gap functions are invariant under all the point group operations, $g=\sigma_x$, $\sigma_y$, or $\sigma_x\sigma_y$:
$$
  \Delta_n(g^{-1}k_x)=\Delta_n(\mp k_x)=\Delta_n(k_x).
$$
Regarding the TR operation, it follows from Eq. (\ref{c-tilde c}) and the antiunitarity of time reversal that it is equivalent to the complex conjugation, i.e.,
$\Delta_n(k_x)\to\Delta^*_n(k_x)$.
Finally, under an arbitrary rotation of the Bloch state phases, $|k_x,n\rangle\to e^{i\theta_n(k_x)}|k_x,n\rangle$, the gap functions remain invariant. 

In a BCS-type model, the gap functions are nonzero only in the vicinity of the Fermi level, where their $k_x$-dependence can be neglected. Therefore, 
the superconducting state can be described by a set of complex order parameters $\Delta_1,...,\Delta_N$, one per each pair of the Fermi points. 
The bulk quasiparticle spectrum in the $n$th band obtained from Eq. (\ref{H-MF}) 
consists of two electron-hole symmetric branches, $\pm\sqrt{\xi_n^2+|\Delta_n|^2}$, with the energy gap equal to $|\Delta_n|$. The stable uniform superconducting states are found by minimizing the Ginzburg-Landau free energy. 
In addition to the TR invariant states, in which the phases of the gap functions are either $0$ or $\pi$, there is also a phenomenological possibility of TR symmetry-breaking states.\cite{Sam15}

\section{Fermionic boundary modes}
\label{sec: ABS}

While the Bogoliubov quasiparticles in the bulk are gapped, there might exist subgap states bound to the ends of the wire.
Let us consider a half-infinite superconductor at $x\geq 0$. To make analytical progress, we neglect self-consistency and assume that the order parameters $\Delta_1,...,\Delta_N$ do not depend on $x$. 
The quasiparticle wave function in the $n$th band is an electron-hole spinor, which can be represented in the semiclassical, or Andreev,\cite{And64} approximation
as $e^{irk_{F,n}x}\psi_{n,r}(x)$, where $r=\pm$ characterizes the direction of the Fermi wave vector $rk_{F,n}$ (which is not the same as the direction of propagation, see below). 
The Andreev envelope function $\psi_{n,r}$ varies slowly on the scale of the Fermi wavelength $k_{F,n}^{-1}$ and satisfies the following equation:    
\begin{equation}
\label{And-eq-gen}
	\left(\begin{array}{cc}
		-iv_{n,r}\dfrac{d}{dx} & \Delta_n \\
		\Delta^*_n & iv_{n,r}\dfrac{d}{dx}
	\end{array}\right)\psi_{n,r}=E\psi_{n,r}.
\end{equation}
Here $v_{n,r}=(\partial\xi_n/\partial k_x)|_{k_x=rk_{F,n}}$ is the group velocity and $\Delta_n\equiv\Delta_n(+k_{F,n})=\Delta_n(-k_{F,n})$ 
is the gap function affecting the quasiparticles near the Fermi point $rk_{F,n}$ in the $n$th band. Note that $v_{n,-}=-v_{n,+}$ and $|v_{n,\pm}|\equiv v_{F,n}$. 

We focus on the ABS localized near $x=0$. The corresponding solution of Eq. (\ref{And-eq-gen}) has the form 
$\psi_{n,r}(x)=\phi(rk_{F,n})e^{-\Omega_nx/|v_{n,r}|}$, where 
\begin{eqnarray}
\label{Andreev amplitude}
	\phi(rk_{F,n}) &\equiv& \psi_{n,r}(x=0)\nonumber\\
	&=& C(rk_{F,n})\left(\begin{array}{c}
		\dfrac{\Delta_n}{E-i\Omega_n\sgn v_{n,r}} \\ 1
	\end{array}\right),
\end{eqnarray}
$\Omega_n=\sqrt{|\Delta_n|^2-E^2}$, and $C$ is a coefficient. 
The Andreev approximation breaks down near the boundary of the system due to a rapid variation of the lattice potential, which causes elastic transitions between the states corresponding to different $rk_{F,n}$. 
The general solution for the Bogoliubov quasiparticle wave function in the bulk is given by a linear combination of $2N$ semiclassical expressions corresponding to all possible Fermi wave vectors:
\begin{equation}
\label{Psi-ABS}
  \Psi(x)=\sum_{n=1}^N\sum_{r=\pm}\phi(rk_{F,n})e^{irk_{F,n}x}e^{-\Omega_n x/|v_{n,r}|}.
\end{equation}
The bound state energy, which has to be inside the bulk gaps, i.e., $|E|<|\Delta_n|$ for all $n$, is found using the boundary conditions at $x=0$. 

Depending on the sign of the group velocity, the Fermi wave vectors are classified as either incident, for which $v_{n,r}<0$, or reflected, for which $v_{n,r}>0$. 
We denote the former $k^{\mathrm{in}}_{1},...,k^{\mathrm{in}}_N$ and the latter $k^{\mathrm{out}}_{1},...,k^{\mathrm{out}}_N$, so that $k^{\mathrm{in}}_n=-k_{F,n}\sgn v_{n,+}$ and $k^{\mathrm{out}}_n=k_{F,n}\sgn v_{n,+}$,
see Fig. \ref{fig: Fermi points}. 
According to Ref. \onlinecite{Shel-bc}, the Andreev amplitudes at $x=0$ for the reflected waves are related to those for the incident waves as follows:
\begin{equation}
\label{Shelankov-bc}
  \phi(k^{\mathrm{out}}_n)=\sum_{m=1}^N S_{nm}\phi(k^{\mathrm{in}}_{m}).
\end{equation}
The coefficients $S_{nm}$ form a unitary $N\times N$ matrix ($S$ matrix), which is an electron-hole scalar, determined by the microscopic details of the boundary scattering at the Fermi level in the normal state. 
The Andreev amplitudes in Eq. (\ref{Shelankov-bc}) are given by
\begin{equation}
\label{phi-in-out}
  \phi(k^{\mathrm{in}}_n)=A_n\left(\begin{array}{c}
		\alpha^{\mathrm{in}}_n \\ 1
	\end{array}\right),\quad
  \phi(k^{\mathrm{out}}_n)=B_n\left(\begin{array}{c}
		\alpha^{\mathrm{out}}_n \\ 1
	\end{array}\right),
\end{equation}
with $A_n=C(k^{\mathrm{in}}_n)$, $B_n=C(k^{\mathrm{out}}_n)$, and
$$
  \alpha^{\mathrm{in(out)}}_n=\frac{\Delta_n}{E\pm i\sqrt{|\Delta_n|^2-E^2}}.
$$
Inserting Eq. (\ref{phi-in-out}) into the boundary conditions (\ref{Shelankov-bc}), we arrive at the following equation for the ABS energy:
\begin{equation}
\label{ABS-energy-equation}
  \det\hat W(E)=0,
\end{equation}
where
\begin{equation}
\label{W-def}
  \hat W(E)=\hat S-\hat M_{\mathrm{out}}^\dagger(E)\hat S\hat M_{\mathrm{in}}(E),
\end{equation}
$\hat M_{\mathrm{in}}=\diag(\alpha^{\mathrm{in}}_1,...,\alpha^{\mathrm{in}}_N)$, and $\hat M_{\mathrm{out}}=\diag(\alpha^{\mathrm{out}}_1,...,\alpha^{\mathrm{out}}_N)$. The degeneracy of the ABS energy $E$ 
is equal to the dimension of the kernel of the $N\times N$ matrix $\hat W(E)$.

The surface ABS have been extensively studied in other superconductors, such as high-$T_c$ cuprates.\cite{ZBCP} Domain walls separating two degenerate 
superconducting states, e.g., $k_x+ik_y$ and $k_x-ik_y$ chiral $p$-wave states, can also trap subgap quasiparticles in their vicinity.\cite{SC-DWs} For a recent study of the ABS in
2D noncentrosymmetric superconductors, see Ref. \onlinecite{SM16}.
The ABS origin in all these systems can be attributed to the gap function's variation along the quasiparticle's semiclassical 
trajectory. For example, the $d$-wave gap changes sign upon a specular reflection from a suitably oriented surface, leading to a dispersionless zero-energy surface ABS.\cite{Hu94}  
By analogy, if in our case the gap functions ``seen'' by the quasiparticles before and after the boundary reflection are different, then one can expect the ABS zero modes.

\section{TR invariant superconducting state}
\label{sec: TRI}

In the rest of the paper we focus on the TR invariant superconducting states, in which the gap functions can be chosen to be real, with the phases taking just two values, either $0$ or $\pi$. 
In this case, one has 
$$
  \alpha^{\mathrm{in}}_n=\alpha^{\mathrm{out},*}_n=\alpha_n(E)=\frac{\Delta_n}{E+i\sqrt{\Delta_n^2-E^2}}.
$$
Without any loss of generality we assume that the first $N_+$ bands have positive gap functions, while the remaining $N_-=N-N_+$ bands have negative gap functions. It is sufficient to consider 
$0\leq N_+\leq N/2$ (recall that $N$ is even), because at $N/2<N_+\leq N$ one can flip the signs of all gap functions by a gauge transformation.
The scattering matrix can be represented in the block form as follows:
\begin{equation}
\label{S-R}
  \hat S=\left(\begin{array}{cc}
          \hat R_{++} & \hat R_{+-} \\
          \hat R_{-+} & \hat R_{--}  \\ 
         \end{array}\right),
\end{equation}
where $\hat R_{ss'}$ is a $N_{s}\times N_{s'}$ matrix ($s,s'=\pm$). From Eq. (\ref{W-def}) we obtain:
\begin{equation}
\label{W-TRI}
  \hat W(E)=\left(\begin{array}{cc}
          \hat R_{++}-\hat m_+\hat R_{++}\hat m_+ & \hat R_{+-}-\hat m_+\hat R_{+-}\hat m_- \\
          \hat R_{-+}-\hat m_-\hat R_{-+}\hat m_+ & \hat R_{--}-\hat m_-\hat R_{--}\hat m_-  \\ 
         \end{array}\right),
\end{equation}
where $\hat m_+(E)=\diag(\alpha_1,...,\alpha_{N_+})$ and $\hat m_-(E)=\diag(\alpha_{N_++1},...,\alpha_N)$. The solution of the ABS energy equation (\ref{ABS-energy-equation}) depends on the 
microscopic details of the system, namely, on the gap magnitudes and the $S$ matrix elements. However, if one focuses on the solutions with zero energy, then it is possible to make considerable progress and 
get some universal results.  

To find the number ${\cal N}_0$ of the ABS zero modes, we put $E=0$ in Eq. (\ref{W-TRI}), use $\hat m_+(0)=-i\mathbb{1}_{N_+\times N_+}$ and $\hat m_-(0)=i\mathbb{1}_{N_-\times N_-}$, and obtain:
\begin{equation}
\label{W0}
  \hat W(0)=2\left(\begin{array}{cc}
          \hat R_{++} & 0 \\
          0 & \hat R_{--}  \\ 
         \end{array}\right).
\end{equation}
Therefore,
\begin{equation}
\label{number-zero-modes}
  {\cal N}_0=\dim(\ker\hat R_{++})+\dim(\ker\hat R_{--}).
\end{equation}
In general, since there is no reason for the blocks of the $S$ matrix to have zero eigenvalues, we have $\dim(\ker\hat R_{++})=\dim(\ker\hat R_{--})=0$ and ${\cal N}_0=0$. 
However, there are some cases in which zero eigenvalues are required by the TR symmetry and the unitarity of the scattering matrix.  

In order to understand the constraints imposed on the $S$ matrix by the TR invariance of the normal state, 
we express the wave function of normal electrons away from the boundary as a superposition of $N$ incident and $N$ reflected states:
\begin{equation}
\label{Psi-general-N}
  |\Psi\rangle=\sum_{n=1}^N\left(A_n|k^{\mathrm{in}}_n\rangle+B_n|k^{\mathrm{out}}_n\rangle\right),
\end{equation}
where $|k\rangle\equiv|k,n\rangle$ is a shorthand notation for the spinor Bloch state corresponding to the Fermi wave vector $k=k^{\mathrm{in}}_n$ or $k^{\mathrm{out}}_n$. 
The scattering matrix is defined by 
\begin{equation}
\label{S-matrix-N}
  B_n=\sum_m S_{nm}A_m.
\end{equation}
Applying the TR operation to the wave function (\ref{Psi-general-N}) and using Eq. (\ref{t_n}), we obtain: 
\begin{eqnarray*}
 K|\Psi\rangle &=& \sum_n \left[A_n^*t_n(k^{\mathrm{in}}_n)|-k^{\mathrm{in}}_n\rangle+B_n^*t_n(k^{\mathrm{out}}_n)|-k^{\mathrm{out}}_n\rangle\right]\\
 &=& \sum_n \left[B_n^*t_n(k^{\mathrm{out}}_n)|k^{\mathrm{in}}_n\rangle+A_n^*t_n(k^{\mathrm{in}}_n)|k^{\mathrm{out}}_n\rangle\right].
\end{eqnarray*}
Note how the ``in''- and ``out''-states are interchanged by time reversal. If the bulk Hamiltonian is TR invariant and the boundary is nonmagnetic, then one can expect the same 
$S$-matrix relations between the incident and reflected states in $|\Psi\rangle$ and $K|\Psi\rangle$, therefore
\begin{equation}
\label{TR-S-matrix-N}
  A_n^*t_n(k^{\mathrm{in}}_n)=\sum_m S_{nm}B_m^*t_m(k^{\mathrm{out}}_{m}).
\end{equation}
Comparing Eqs. (\ref{S-matrix-N}) and (\ref{TR-S-matrix-N}) and taking into account the $S$-matrix unitarity, we obtain:
\begin{equation}
\label{TR-S}
  S_{mn}=t_n^*(k^{\mathrm{in}}_n)S_{nm}t_m(k^{\mathrm{out}}_{m}).
\end{equation}
The TR phase factors depend on the phase choice for the Bloch states, see Eq. (\ref{t-transformation}). In particular, one can make $t_n(k_x)$ real and equal to $+1$ for $k_x=k^{\mathrm{out}}_{n}$, and to $-1$ for 
$k_x=k^{\mathrm{in}}_{n}=-k^{\mathrm{out}}_n$. Then it follows from Eq. (\ref{TR-S}) that in this basis the $S$ matrix is antisymmetric:
\begin{equation}
\label{S-antisymmetric}
  S_{mn}=-S_{nm}.
\end{equation}
This means, in particular, that $S_{nn}=0$, i.e., the backscattering of quasiparticles into the same band is forbidden by the TR symmetry. This last conclusion is also confirmed by the dimensional reduction of the 
$S$ matrix for the 2D Rashba model.\cite{KS10,SM16} 

Returning now to counting the zero-energy ABS in the superconducting state, see Eq. (\ref{number-zero-modes}), and using the property (\ref{S-antisymmetric}), we have
$$
  \hat R_{++}=\left(\begin{array}{cccc}
         0 & S_{12} & \ldots & S_{1,N_+} \\
	 \vdots & \vdots & \ddots & \vdots \\
	 -S_{1,N_+} & -S_{2,N_+} & \ldots & 0
        \end{array}\right)
$$ 
and
$$
  \hat R_{--}=\left(\begin{array}{cccc}
         0 & S_{N_++1,N_++2} & \ldots & S_{N_++1,N} \\
	 \vdots & \vdots & \ddots & \vdots \\
	 -S_{N_++1,N} & -S_{N_++2,N} & \ldots & 0
        \end{array}\right).
$$
If $N_+=1$, then $\hat R_{++}=0$ and, therefore, $\dim(\ker\hat R_{++})=1$. In this case, $\hat R_{--}$ also has a zero eigenvalue. 
Indeed, from the unitarity of the scattering matrix (\ref{S-R}) we obtain:
$\hat R_{+-}\hat R_{+-}^\dagger=1$ and $\hat R_{--}\hat R_{+-}^\dagger=\bm{0}$. The column vector 
\begin{equation}
\label{Rpm}
  \hat R_{+-}^\dagger=\left(\begin{array}{c}
                                    S_{12}^*\\
				    \vdots \\
				    S_{1N}^*
                                   \end{array}\right)
\end{equation}
has unit norm and is a zero mode of $\hat R_{--}$. Therefore, if $N_+=1$, then $\dim(\ker\hat R_{++})=\dim(\ker\hat R_{--})=1$ and ${\cal N}_0=2$.

If $N_+$ is an even number, then zero modes are not required by the unitarity and TR invariance of the $S$ matrix. However, if $N_+$ is odd, then both $\hat R_{++}$ and
$\hat R_{--}$ are odd-dimensional antisymmetric matrices and, therefore, have at least one zero eigenvalue each. Thus, we come to the conclusion that the non-accidental zero-energy ABS exist only if the gap sign
in an odd number of bands is different from all other bands. Namely,
\begin{equation}
\label{zero-modes-final}
  {\cal N}_0=\left\{\begin{array}{ll}
                    2,\quad & \mathrm{if}\ N_+\ \mathrm{odd},\\
		    0,\quad & \mathrm{if}\ N_+\ \mathrm{even}. 
                    \end{array}\right.
\end{equation}
The zero-mode wave functions for $N_+=1$ are discussed in Sec. \ref{sec: ABS wave functions}. 

In the case of just two nondegenerate bands with $N_+=N_-=1$, Eq. (\ref{zero-modes-final}) agrees with the dimensional reduction of the known result for a 2D Rashba superconductor in the $x>0$ half-plane. 
The latter is in a topologically nontrivial state if the triplet pairing channel dominates the singlet one. In the band representation, this corresponds to the gap functions having opposite signs 
in the two helicity bands. 
Then, according to Refs. \onlinecite{SF09} and \onlinecite{TYBN09}, there exist two helical ABS modes with a linear dispersion, which counterpropagate along the surface. 
The 1D Rashba superconductor is obtained by looking only at the normally incident quasiparticles. Since both helical modes have zero energy at $k_y=0$, we have ${\cal N}_0=2$.   

The result (\ref{zero-modes-final}) is also consistent with the topological arguments. The bulk-boundary correspondence principle states that the number of the boundary zero modes is related 
to a topological invariant in the bulk.\cite{Volovik-book,top-SC} Our system belongs to the symmetry class DIII, which can be characterized in 1D by a $\mathbb{Z}_2$ invariant.\cite{RSFL10}  
According to Ref. \onlinecite{QHZ10}, this invariant has the following form:
\begin{equation}
\label{Z2-invariant}
  \prod_{n=1}^N\sgn\Delta_n=(-1)^{N_+}.
\end{equation}
The states with $N_+$ odd are $\mathbb{Z}_2$-nontrivial and should have a pair of the ABS zero modes, in agreement with Eq. (\ref{zero-modes-final}).

\subsection{Zero-mode wave function}
\label{sec: ABS wave functions}

The wave function of the ABS zero modes is obtained from Eqs. (\ref{Psi-ABS}) and (\ref{phi-in-out}):
\begin{eqnarray}
\label{Psi-ZEABS}
  \Psi_0(x)=\sum_{n=1}^N\left[ A_n\left(\begin{array}{c}
                                                           -i\sgn\Delta_n \\ 1
                                                          \end{array}\right)e^{ik^{\mathrm{in}}_nx} \right.\nonumber\\
  +\left. B_n\left(\begin{array}{c}
                                                          i\sgn\Delta_n \\ 1
                                                          \end{array}\right)e^{ik^{\mathrm{out}}_nx} \right]e^{-\kappa_nx},
\end{eqnarray}
where $\kappa_n=|\Delta_n|/v_{F,n}$. It follows from the boundary conditions (\ref{Shelankov-bc}) that the amplitudes of the reflected states are given by $B_n=\sum_m S_{nm}A_m$, while the amplitudes 
of the incident states satisfy two decoupled equations, separately in the positive-gap and the negative-gap bands: 
\begin{equation}
\label{RCRC}
  \hat R_{++}\bm{A}_+=\bm{0},\quad \hat R_{--}\bm{A}_-=\bm{0},
\end{equation}
where $\bm{A}_+=(A_1,...,A_{N_+})^\top$ and $\bm{A}_-=(A_{N_++1},...,A_{N})^\top$.

For $N_+=1$, we have $\hat R_{++}=0$ and the first equation in Eq. (\ref{RCRC}) is solved by a constant $C_1$, with the amplitudes of the reflected states 
given by $B_n=C_1S_{n1}$. It follows from Eq. (\ref{Rpm}) that the solution of the second equation in Eq. (\ref{RCRC}) has the form $A_{-,n}=C_2S^*_{1n}$, where $C_2$ is a constant, and
the amplitudes of the corresponding reflected states are given by $B_n=C_2\sum_m S_{nm}S^*_{1m}=C_2\delta_{n1}$. Putting everything together, we obtain the following wave function:
\begin{eqnarray}
\label{Psi0-final}
  && \Psi_0(x)=C_1\left(\begin{array}{c}
               -i \\ 1
            \end{array}\right)\nonumber\\
  &&  \times\left(e^{ik^{\mathrm{in}}_1x}e^{-\kappa_1x}+\sum_{n=2}^NS_{n1}e^{ik^{\mathrm{out}}_nx}e^{-\kappa_nx}\right)\nonumber\\
  && +C_2\left(\begin{array}{c}
               i \\ 1
            \end{array}\right)\nonumber\\
  &&  \times\left(\sum_{n=2}^NS^*_{1n}e^{ik^{\mathrm{in}}_nx}e^{-\kappa_nx}+e^{ik^{\mathrm{out}}_1x}e^{-\kappa_1x}\right),
\end{eqnarray}
which describes a twofold degenerate zero-energy state.

Using Eq. (\ref{Psi0-final}), one arrives at the following simple qualitative picture of the origin of the ABS zero modes in the $N_+=1$ case. The $C_1$ term describes
an incident quasiparticle in the $n=1$ band, ``seeing'' a positive gap, which is reflected into one of the $n>1$ bands, all of which have negative gaps. The $C_2$ term describes incident quasiparticles in the negative-gap bands,
which are reflected into the positive-gap band. In both processes, the gap function along the quasiparticle trajectory 
necessarily changes sign, leading to the zero-energy ABS.

\section{Conclusions}
\label{sec: Conclusion}

Superconducting quantum wires are intrinsically noncentrosymmetric, due to the breaking of the up-down reflection symmetry by the substrate. The antisymmetric SO coupling splits the Bloch bands almost everywhere, 
except the center and the boundaries of the 1D Brillouin zone. The form (``direction'') of the SO coupling depends on the presence or absence of additional mirror reflection symmetries of the quasi-1D confining potential. 
In a TR invariant normal state, the 1D Fermi surface consists of an even number $N$ of symmetrical Fermi-point pairs. 
If the SO band splitting is sufficiently large, then the interband Cooper pairing is suppressed and the superconductivity is characterized by $N$ intraband
gap functions $\Delta_n$. These gap functions describe the pairing of quasiparticles in the time-reversed states in the same band, are even in momentum and invariant under all point group operations. 

Since superconductivity is a Fermi-surface phenomenon, one can focus on the quasiparticle properties near the Fermi points and use the semiclassical formalism based on the Andreev equations and the $S$-matrix 
boundary conditions. In this way, one can make a considerable progress in the general multiband case and obtain some model-independent results.
Assuming a TR invariant superconducting state, we found that there is a pair of the zero-energy Andreev modes (Majorana quasiparticles) 
at the end of the wire only if the gap sign in an odd number of bands is different from all other bands.

\acknowledgments
This work was supported by a Discovery Grant from the Natural Sciences and Engineering Research Council of Canada. The author is grateful to NORDITA, Stockholm for hospitality during 
the workshop ``Multi-Component and Strongly-Correlated Superconductors 2016'', and also to the anonymous referee, who pointed out a mistake in the first version of the manuscript.

\appendix

\section{Inversion-asymmetric SOC in 1D}
\label{sec: ASOC-1D}

Adapting the results of Ref. \onlinecite{Sam09} to the 1D case, we arrive at the general form of the Hamiltonian of noninteracting electrons in a TR invariant 1D crystal lattice with the SO coupling:
\begin{eqnarray}
\label{H pseudospin}
    \hat H_0=\sum_{k_x,ab}\sum_{\alpha,\beta}[\epsilon_a(k_x)\delta_{ab}\delta_{\alpha\beta}+
    iA_{ab}(k_x)\delta_{\alpha\beta}\nonumber\\
    +\bm{B}_{ab}(k_x)\bm{\sigma}_{\alpha\beta}]\hat a^\dagger_{k_xa\alpha}\hat a_{k_xb\beta}.
\end{eqnarray}
Here the indices $a,b$ label the twofold degenerate bands with the dispersion $\epsilon_a(k_x)=\epsilon_a(-k_x)$ obtained from the inversion-symmetric parts of the lattice potential and the SO coupling, 
$\alpha,\beta=\uparrow,\downarrow$ denote the pseudospin, $\hat{\bm{\sigma}}=(\hat\sigma_1,\hat\sigma_2,\hat\sigma_3)$ are the Pauli matrices, and the wave-vector summation is performed over the 1D BZ. 
All effects of the inversion-antisymmetric parts of the potential and the SO coupling are contained in the last two terms, which are odd in $k_x$. 
It follows from the requirements of Hermiticity, TR invariance, and the lattice periodicity that $A_{ab}$ and $\bm{B}_{ab}$ are real, periodic in the reciprocal space, and 
satisfy $A_{ab}(k_x)=-A_{ba}(k_x)$ and $\bm{B}_{ab}(k_x)=\bm{B}_{ba}(k_x)$. The 1D point group might impose additional constraints, see Sec. \ref{sec: electron bands}.

Diagonalizing the Hamiltonian (\ref{H pseudospin}) one obtains the Bloch bands, denoted by $\xi_n(k_x)$, which are even in $k_x$ and 
remain twofold degenerate only at some isolated points in the BZ, where $\bm{B}_{ab}$ vanishes. 
It is easy to see that this happens generically only at the TR invariant wave vectors $k_x=K$, satisfying $-K=K+G$, where $G=\pm 2\pi/d$ is the 1D reciprocal lattice vector. 
There are two TR invariant points in the BZ, given by $K_1=0$ and $K_2=\pi/d$. At these points, we have $\bm{B}_{ab}(K)=-\bm{B}_{ab}(-K)=-\bm{B}_{ab}(K+G)=-\bm{B}_{ab}(K)$, therefore 
$\bm{B}_{ab}(K_{1,2})=\bm{0}$. 
For the same reason, $A_{ab}(K_{1,2})=0$. 

\begin{figure}
\includegraphics[width=7cm]{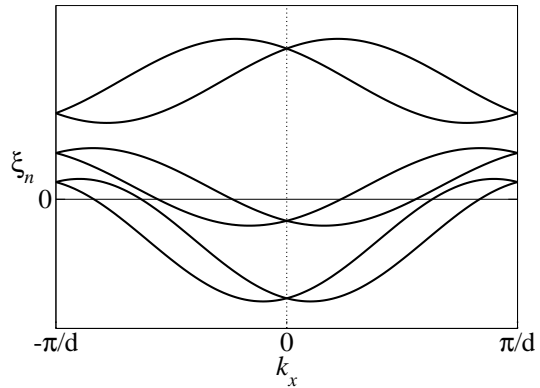}
\caption{The 1D Bloch bands split by the antisymmetric SO coupling. The bands remain twofold degenerate at the TR invariant points $k_x=0$ and $\pi/d$. The chemical potential shown
corresponds to $N=4$.}
\label{fig: bands}
\end{figure}

Any band crossings at $k_x\neq K_{1,2}$ are accidental and can, therefore, be removed by a small variation of the system parameters.
Thus, the diagonalization of Eq. (\ref{H pseudospin}) produces nondegenerate bands that come in pairs connected at the center and the boundaries of the 1D BZ, 
as shown in Fig. \ref{fig: bands}. Keeping just one such pair of bands, corresponding to $a=0$, using the fact that $A_{00}(k_x)=0$, and introducing the notation $\bm{B}_{00}(k_x)=\bgam(k_x)$,
one arrives at the Rashba model, see Eq. (\ref{H-Rashba}).

The Fermi points in the $n$th band are given by the roots of the equation $\xi_n(k_x)=0$, see Fig. \ref{fig: bands}. These roots always come in pairs, $\pm k_{F,n}$. It is easy to see that, barring some exceptional values 
of the chemical potential, at which either $\xi_n(0)=0$ or $\xi_n(\pi/d)=0$, the total number of the pairs of the Fermi points is even,
see also Ref. \onlinecite{WBZ06}. This last statement is no longer true if the TR symmetry is broken in the normal state,
e.g., by an external magnetic field, which can lift the band degeneracies at the TR invariant momenta.

\newpage



\begin{thebibliography}{999}

\bibitem{MFs-review} 
J. Alicea, Rep. Prog. Phys. \textbf{75}, 076501 (2012); M. Leijnse and K. Flensberg, Semicond. Sci. Technol. \textbf{27}, 124003 (2012).

\bibitem{top-SC}
B. A. Bernevig, \textit{Topological Insulators and Topological Superconductors} (Princeton University Press, USA, 2013).

\bibitem{Kit01}
A. Y. Kitaev, Physics-Uspekhi \textbf{44}, 131 (2001).  

\bibitem{TRB-1D}
R. M. Lutchyn, J. D. Sau, and S. Das Sarma, Phys. Rev. Lett. \textbf{105}, 077001 (2010); Y. Oreg, G. Refael, and F. von Oppen, Phys. Rev. Lett. \textbf{105}, 177002 (2010).

\bibitem{InSb-wire-1}
V. Mourik, K. Zuo, S. M. Frolov, S. R. Plissard, E. P. A. M. Bakkers, and L. P. Kouwenhoven, Science \textbf{336}, 1003 (2012). 

\bibitem{InSb-wire-2}
H. O. H. Churchill, V. Fatemi, K. Grove-Rasmussen, M. T. Deng, P. Caroff, H. Q. Xu, and C. M. Marcus, Phys. Rev. B \textbf{87}, 241401(R) (2013).

\bibitem{Fe-chain}
S. Nadj-Perge, I. K. Drozdov, J. Li, H. Chen, S. Jeon, J. Seo, A. H. MacDonald, B. A. Bernevig, and A. Yazdani, Science \textbf{346}, 602 (2014);
R. Pawlak, M. Kisiel, J. Klinovaja, T. Meier, S. Kawai, T. Glatzel, D. Loss, and E. Meyer, arXiv:1505.06078. 

\bibitem{TRI-1D}
C. L. M. Wong and K. T. Law, Phys. Rev. B 86, 184516 (2012); F. Zhang, C. L. Kane, and E. J. Mele, Phys. Rev. Lett. \textbf{111}, 056402 (2013); 
A. Keselman, L. Fu, A. Stern, and E. Berg, Phys. Rev. Lett. \textbf{111}, 116402 (2013); E. Gaidamauskas, J. Paaske, and K. Flensberg,
Phys. Rev. Lett. \textbf{112}, 126402 (2014); A. Haim, A. Keselman, E. Berg, and Y. Oreg, Phys. Rev. B \textbf{89}, 220504(R) (2014); A. Haim, E. Berg, K. Flensberg, and Y. Oreg, Phys. Rev. B \textbf{94}, 161110(R) (2016).

\bibitem{Rashba-model}
Yu. A. Bychkov and E. I. Rashba,  Pis'ma Zh. Eksp. Teor. Fiz. \textbf{39}, 66 (1984) [JETP Lett. \textbf{39}, 78 (1984)].

\bibitem{Manchon15}
A. Manchon, H. C. Koo, J. Nitta, S. M. Frolov, and R. A. Duine, Nature Materials \textbf{14}, 871 (2015).

\bibitem{NCSC-book}
\textit{Non-centrosymmetric Superconductors: Introduction and Overview}, ed. by E. Bauer and M. Sigrist, Lecture Notes in Physics \textbf{847} (Springer, Heidelberg, 2012). 

\bibitem{Kneid15}
F. Kneidinger, E. Bauer, I. Zeiringer, P. Rogl, C. Blaas-Schenner, D. Reith, and R. Podloucky, Physica C \textbf{514}, 388 (2015).

\bibitem{Smid16}
M. Smidman, M. B. Salamon, H. Q. Yuan, and D. F. Agterberg, arXiv:1609.05953.

\bibitem{Sam15}
K. V. Samokhin, Ann. Phys. (N.Y.) \textbf{359}, 385 (2015); K. V. Samokhin, Phys. Rev. B \textbf{92}, 174517 (2015).

\bibitem{1D}
Note that a ``pure'' 1D system is a mathematical idealization, described by a potential which depends only on $x$. In this case, there are just two point groups: $\mathbf{C}_1=\{E\}$ and $\mathbf{D}_x=\{E,\sigma_x\}$.

\bibitem{t-factor}
L. P. Gor'kov and E. I. Rashba, Phys. Rev. Lett. \textbf{87}, 037004 (2001);
I. A. Sergienko and S. H. Curnoe, Phys. Rev. B \textbf{70}, 214510 (2004); K. V. Samokhin, Phys. Rev. B \textbf{70}, 104521 (2004). 

\bibitem{And64}
A. F. Andreev, Zh. Eksp. Teor. Fiz. \textbf{46}, 1823 (1964) [Sov. Phys.-JETP \textbf{19}, 1228 (1964)]; Ch. Bruder, Phys. Rev. B \textbf{41}, 4017 (1990); I. Adagideli, P. M. Goldbart, A. Shnirman, and A. Yazdani, 
Phys. Rev. Lett. \textbf{83}, 5571 (1999).

\bibitem{Shel-bc}
A. L. Shelankov, Pis'ma Zh. Eksp. Teor. Fiz. \textbf{32}, 122 (1980) [JETP Lett. \textbf{32}, 111 (1980)]; A. Shelankov and M. Ozana, Phys. Rev. B \textbf{61}, 7077 (2000).

\bibitem{ZBCP} 
S. Kashiwaya and Y. Tanaka, Rep. Prog. Phys. \textbf{63}, 1641 (2000).

\bibitem{SC-DWs} 
T. L. Ho, J. R. Fulco, J. R. Schrieffer, and F. Wilczek, Phys. Rev. Lett. \textbf{52}, 1524 (1984); M. Matsumoto and M. Sigrist, J. Phys. Soc. Jpn. \textbf{68}, 994 (1999). 

\bibitem{SM16}
K. V. Samokhin and S. P. Mukherjee, Phys. Rev. B \textbf{94}, 104523 (2016).

\bibitem{Hu94}
C.-R. Hu, Phys. Rev. Lett. \textbf{72}, 1526 (1994).

\bibitem{KS10}
A. Khaetskii and E. Sukhorukov, JETP Letters \textbf{92}, 244 (2010).

\bibitem{SF09}
M. Sato and S. Fujimoto, Phys. Rev. B \textbf{79}, 094504 (2009).

\bibitem{TYBN09}
Y. Tanaka, T. Yokoyama, A. V. Balatsky, and N. Nagaosa, Phys. Rev. B \textbf{79}, 060505(R) (2009).

\bibitem{Volovik-book}
G. E. Volovik, \textit{The Universe in a Helium Droplet} (Clarendon Press, Oxford, 2003).

\bibitem{RSFL10}
S. Ryu, A. P. Schnyder, A. Furusaki, and A. W. W. Ludwig, New J. Phys. \textbf{12}, 065010 (2010).

\bibitem{QHZ10}
X.-L. Qi, T. L. Hughes, and S.-C. Zhang, Phys. Rev. B \textbf{81}, 134508 (2010).

\bibitem{Sam09}
K. V. Samokhin, Ann. Phys. (N. Y.) \textbf{324}, 2385 (2009).

\bibitem{WBZ06}
C. Wu, B. A. Bernevig, and S.-C. Zhang, Phys. Rev. Lett. \textbf{9}6, 106401 (2006).

\end{thebibliography}
\end{document}